\documentclass{elsart}
\usepackage[numbers,sort&compress]{natbib}
\usepackage{color,setspace,times}
\usepackage{amssymb,natbib,amsmath,graphicx,color,rotating,subfigure,url}
\usepackage{lineno}

\journal{Physica A}

\begin{document}

\begin{frontmatter}

\title{Geography and distance effect on financial dynamics in the Chinese stock market}

\author[SE]{Xing Li},
\author[SE]{Tian Qiu \corauthref{cor}},
\corauth[cor]{Corresponding author. Address: 696 South Fenghe
Avenue, School of Information Engineering, Nanchang Hangkong
University, Nanchang, 330063, China.} \ead{tianqiu.edu@gmail.com}
\author[SE]{Guang Chen},
\author[SA1,SA2]{Li-Xin Zhong},
\author[SA3]{Xiong-Fei Jiang}

\address[SE]{School of Information Engineering, Nanchang Hangkong University, Nanchang, 330063, China}
\address[SA1]{School of Finance and Coordinated Innovation Center of Wealth Management and Quantitative Investment, Zhejiang University of Finance and Economics, Hangzhou, 310018, China}
\address[SA2]{China Academy of Financial Research, Zhejiang University of Finance and Economics, Hangzhou, 310018, China}
\address[SA3]{School of Information Engineering, Ningbo Dahongying University, Ningbo, 315175, China}

\begin{abstract}
Geography effect is investigated for the Chinese stock market including the Shanghai and Shenzhen stock markets, based on the daily data of individual stocks. The Shanghai city and the Guangdong province can be identified in the stock geographical sector. By investigating a geographical correlation on a geographical parameter, the stock location is found to have an impact on the financial dynamics, except for the financial crisis time of the Shenzhen market. Stock distance effect is further studied, with a crossover behavior observed for the stock distance distribution. The probability of the short distance is much greater than that of the long distance. The average stock correlation is found to weakly decay with the stock distance for the Shanghai stock market, but stays nearly stable for different stock distance for the Shenzhen stock market.

\end{abstract}


\end{frontmatter}

\section{Introduction}


Stock market has an essential function to the country economics, therefore the market evolution has attracted a great interest of scientists from different research fields, such as the economists, mathematicians, and physicists. Among them, a lot of physicists devoted to the study of financial dynamics in the past two decades \cite{man95,gop99,liu99,gab03,gu08,mis10,ren10,zha11,qiu12,pod09,ava14,cur14,ken15,qia15}, and some stylized facts have been revealed from the statistical physics perspective. Scaling behavior of the return and return interval distributions has been observed for different markets \cite{man95,gop99,liu99,gu08,qiu08,yam05,wan08,jun08}. Volatility clustering is found to be universal for most markets \cite{liu99,gia01}. The time correlation and time-spatial correlation is widely studied for the financial markets \cite{son11,men14}. Various models have been proposed to understand the underlying mechanism of financial dynamics\cite{egu00,gu09,che13,qiu10}, and the economic dynamics is also investigated from the experimental perspective \cite{hua15a,hua15b}.

Previous studies have gained abundant characteristic of financial dynamics. However, there are few studies focusing on the the geography effect on financial dynamics, by using the long-term  empirical data of individual stocks. In fact, how the geography affects the finance is an important economic issue, and has received a wide discussion from the economic literature. It has ever been regarded that the rapid development of telecommunication and internet has changed the geography role in finance \cite{cas89}. Economic space becomes no longer important \cite{ohm95a,ohm95b,kob97}. However, a contrast point of view says that the spatial effect is still critical\cite{ber97,zha04}. The geographical information has an essential influence on the pattern of cross-border equity \cite{por05}. The cultural distance is also found to contribute to the transaction cost \cite{agg09}. Lucey et al find a higher country-pair linkage for the smaller cultural distance \cite{luc10}. Up to now, how geography and distance affect the financial dynamics still remains controversial.

In this article, we try to understand the geography effect on the stock market, by applying the random matrix theory, cross-correlation function, etc, based on the daily data of the individual stocks of the Chinese stock market. Our results show that the stock location still has an impact on the financial dynamics. The stock distance is found to only have an impact on the Shanghai stock market, but have no influence on the Shenzhen stock market.

\section*{Datasets and geographical sector}
The datasets are based on the daily data of the individual stocks of the Chinese stock market from the Jan. 1, 2005 to Dec. 31, 2010. Since the market experiences the financial crisis around the year 2008, the data cover three stages of the time before, in, and after the financial crisis. To ensure the stock liquidity, only the stocks whose number of trading days is no less than 150 days are selected. Finally, 778 stocks are chosen in the Shanghai stock market (SH), and 474 stocks are chosen in the Shenzhen stock market(SZ). The stock location is denoted as the headquarter location of the company, and the location of the selected stocks covers all the provinces of China. In the Shanghai stock market, the number of stocks whose company headquarters are located in the Shanghai city is 153, occupying 19.67\% of the Shanghai stock market. And in the Shenzhen stock market, the number of stocks whose company headquarters are located in the Guangdong province is 121, occupying 25.53\% of the Shenzhen stock market. It indicates that the company whose stock is of high liquidity still prefers to set the headquarter in the region around the financial center.

Before investigating the geography effect on the financial dynamics, let us introduce the return and correlation definitions. For a stock $i$, the price return ${R_i}(t')$ of time $t'$ is defined as the Logarithm return of the price $P_i(t')$ over one day,

\begin{equation}
{R_i}(t') = \ln {P_i}(t') - \ln {P_i}(t'-1)
\end{equation}

The normalized return ${r_i}(t')$ of stock $i$ is defined as,

\begin{equation}
{r_i}(t') = \frac{{{R_i}(t') - \langle {R_i}(t')\rangle }}{{{\sigma _i}}}
\end{equation}

where ${\sigma _i} = \sqrt {\langle R_i^2\rangle  - {{\langle {R_i}\rangle }^2}}$. The stock correlation of the price return is defined as,

\begin{equation}
{c_{ij}} = \langle {r_i}{r_j}\rangle
\end{equation}

In the past studies, the economic sectors have been widely investigated based on the random matrix theory \cite{pan07,gop01,she09}. Business sectors can be identified for most mature markets, except for the Chinese stock market. The stocks of the Chinese stock market are identified by the ST and Blue-chip sectors \cite{she09}. To understand how the geography affects the financial dynamics, here we apply the random matrix theory to revealing the geographical sector. Based on the stock correlations of the price returns in Eq. (3), the eigenvalues and eigenvectors of the correlation matrix $\texttt{C}$ are analyzed. By searching for the dominant components in the eigenvectors of the first several largest eigenvalues, the geographical sectors can be identified.

\begin{figure}[htb]
\centering
\includegraphics[width=10cm]{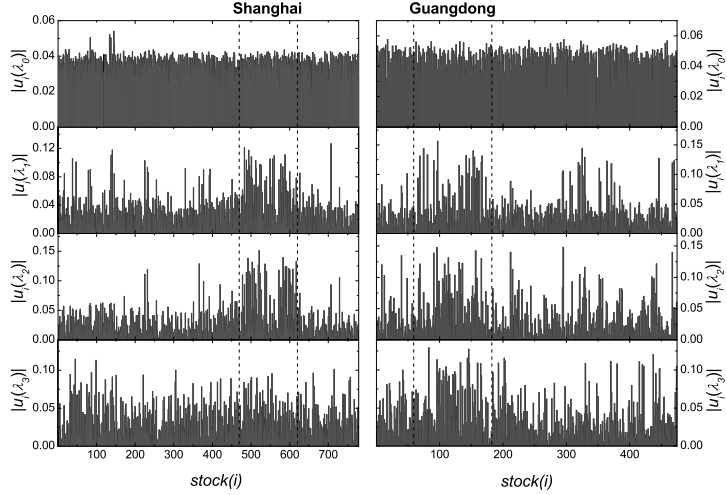}
\caption{The absolute values of the eigenvectors $|u_i|$ are displayed for the first four
largest eigenvalues, with the left panel for the SH, and the right panel for the SZ. The stocks are arranged according to the stock locations, with the order of the location to be the Anhui province, Beijing,  Fujian  province, Gansu province, Guangdong province, Guangxi province, Guizhou province, Hainan province, Hebei province, Henan province, Heilongjiang province, Hubei province, Hunan province, Jilin province, Jiangsu province, Jiangxi province, Liaoning province, Neimenggu,
Ningxia, Qinghai province, Shandong province, Sh$\bar{a}$nxi province, Sh$\check{a}$nxi province,
Shanghai, Sichuan province, Tianjin, Xizhang, Xinjiang, Yunnan province, Zhejiang province, and Chongqing. \label{f4}}
\end{figure}

As shown in Fig. 1, the absolute values $|u_{i}|$ of the eigenvectors are displayed for the first four largest eigenvalues of the Shanghai and Shenzhen stock markets. For both markets, a uniform distribution is observed for the eigenvectors of the largest eigenvalue $\lambda_0$. That is, similarly as the business sector \cite{lal99, ple99}, the largest eigenvalue corresponds to some ''market mode'' for the geographical sector. However, for the second and third largest eigenvalues, it is observed that the stocks located in the Shanghai city dominate the Shanghai stock market, and the stocks located in the Guangdong province dominate the Shenzhen stock market, respectively. For the fourth largest eigenvalue $\lambda_3$, one cannot find the significant component, i.e., no specific sector is identified. The results suggest that the stocks located in the Shanghai city and the Guangdong province play an essential role in the Shanghai and Shenzhen stock market, and the location of financial center is still crucial in financial dynamics.

\begin{figure}[htb]
\centering
\includegraphics[width=12cm]{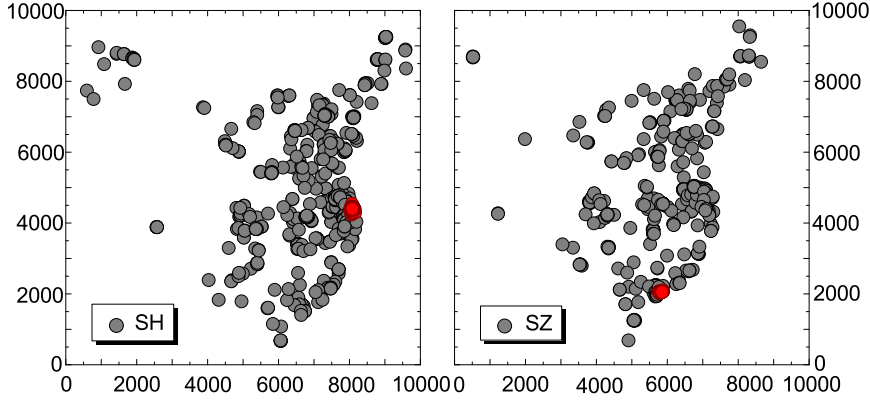}
\caption{An illustration of the stock location distribution is displayed for the SH and SZ markets in the left and right panel, respectively. The X-axis and Y-axis correspond to the longitude and latitude after mapping. The stocks located in the Shanghai and Shenzhen city are marked by the red color. \label{f2}}
\end{figure}

\section*{Geographical correlation dynamics}

Further, the geography effect on the stock correlation dynamics is investigated. we make a two-dimensional map for the stock location. As shown in Fig. 2, all the stocks are mapped onto an $N \times N$ lattice according to the stock location, with $N=10000$. The longitude ($Lon$) of the stock location is mapped onto the $x$-axis of the lattice, and the latitude ($Lat$) of the stock location is mapped onto the $y$-axis of the lattice. The $x$-axis and $y$-axis are uniformly divided by $N$, with the lattice interval of the $x$-axis and $y$-axis to be $l_x=\frac{Max(Lon)-Min(Lon)}{N}$ and $l_y=\frac{Max(Lat)-Min(Lat)}{N}$, respectively. If the longitude of the stock is in the $Min(Lon)+(n-1)l_x \leq Lon < Min(Lon)+nl_x$, and the latitude of the stock is in the $Min(Lat)+(n-1)l_y \leq Lat < Min(Lat)+nl_y$, the stock is then mapped onto the $n_{th}$ lattice, where $n=1,...,N$ is the geographical parameter.

A geographical correlation (GC) is introduced to understand the geographical effect on the stock correlation dynamics, which is defined as,

\begin{equation}
GC(n) = \frac{2}{M(M-1)}\sum\limits_{i,j \in R(n)} {{c_{ij}}}
\end{equation}
where $R(n)$ is the $n \times n$ lattice region, and $M$ is the number of stocks located in the region $R(n)$. That is, the geographical correlation is related to the geographical parameter $n$. Since the financial market experiences a crisis around the year 2008, we investigate the geographical correlation for three stages, i.e., before the financial crisis from Jan. 1, 2005 to Oct. 16, 2007 (BFC), in the financial crisis from Oct. 17, 2007 to Oct. 28, 2008 (IFC), and after the financial crisis from Oct. 29, 2008 to Dec. 31, 2010 (AFC).

The geographical correlation $GC(n)$ on the geographical parameter $n$ is shown in Fig. 3. For the SH market, the $GC(n)$ firstly decays with $n$, and then increases with $n$ as $n$ is greater than about 5000. For the SZ market, the geographical correlation $GC(n)$ increases with $n$ before and after the financial crisis for the $n$ less than about 5500, but decays with $n$ in the financial crisis time. As $n$ is greater than about 5500, the geographical correlation $GC(n)$ then becomes less fluctuated for all the three time periods. The results indicate that the stock location has an impact on the market correlation, especially for the SH market. Moreover, for both markets, it is observed that the correlation of the period in the financial crisis is much higher than that of the period before and after the financial crisis, and the correlation of the period after the financial crisis is also higher than that of the period before the financial crisis. It suggests that, regardless of the location of the company, the stock correlation in the financial crisis period is much stronger than in the normal market period, and the high correlation may relax for a long time.

\begin{figure}[htb]
\centering
\includegraphics[width=10cm]{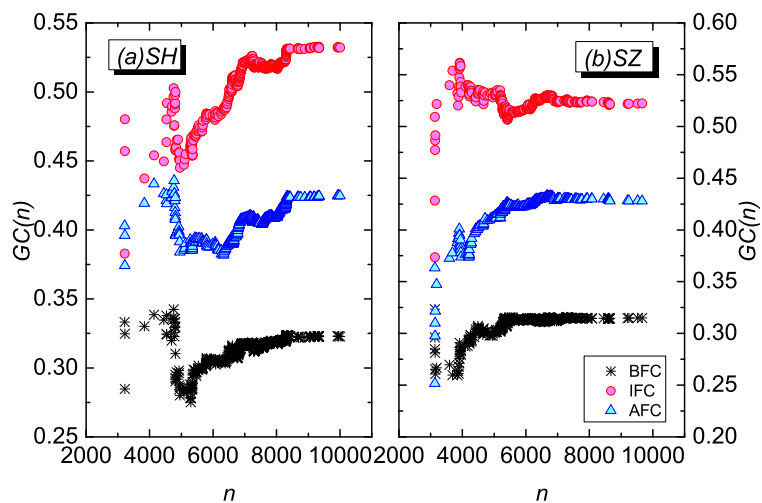}
\caption{The geographical correlation $GC(n)$ on the geographical parameter $n$ is displayed for the SH and SZ. The stars, circles and triangles are for the periods before, in and after the financial crisis, respectively. \label{f3}}
\end{figure}

\begin{figure}[htb]
\centering
\includegraphics[width=10cm]{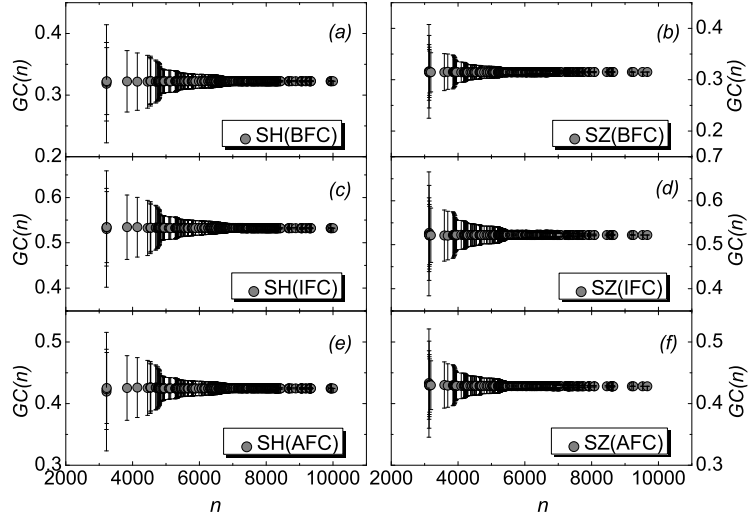}
\caption{The geographical correlation $GC(n)$ of the shuffled data on the geographical parameter $n$ is displayed, with the results being the average over 1000 runs. The panels (a), (c), (e) are for the periods before, in and after the financial crisis of the SH market, and (b), (d), (f) are for the periods before, in and after the financial crisis of the SZ market, respectively. The stems suggest the error bar. \label{f4}}
\end{figure}

To confirm the findings, we shuffle the stocks in different locations, and compute the average geographical correlation over 1000 runs for the shuffled data, as is shown in Fig. 4. Different from the original data, the $GC(n)$ is found to be stable for different geographical parameter $n$. And the stock correlation in the financial crisis time is always higher than that before and after the financial crisis, which is also consistent with that observed in the real market. By performing a t-test for the $GC(n)$ of the real markets and the shuffled data, a significant difference can be detected between them for most time of both markets, with the p-value to be nearly zero. It suggests that the stock location indeed has an impact on the market correlation. An exception is found for the financial crisis period of the SZ market, when the $GC(n)$ of the real market shows no significant difference from that of the shuffled data, with the p-value to be 0.64. As is well known, the Shanghai-listed companies are mainly large enterprises, and the Shenzhen-listed companies are mainly medium-sized and small enterprises. Our results to some extent imply that the stock dynamics of the medium-sized and small enterprises show more homogeneous behavior when encountering the financial crisis, and the location then becomes less important.

\begin{figure}[htb]
\centering
\includegraphics[width=10cm]{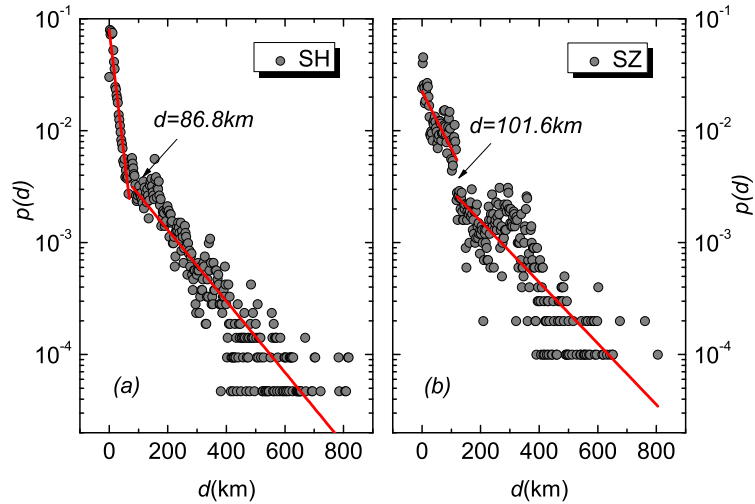}
\caption{The stock distance distributions are displayed for (a) the SH market and (b) the SZ market on a log-linear scale. The solid lines are the exponential fits.\label{f5}}
\end{figure}

\begin{figure}[htb]
\centering
\includegraphics[width=10cm]{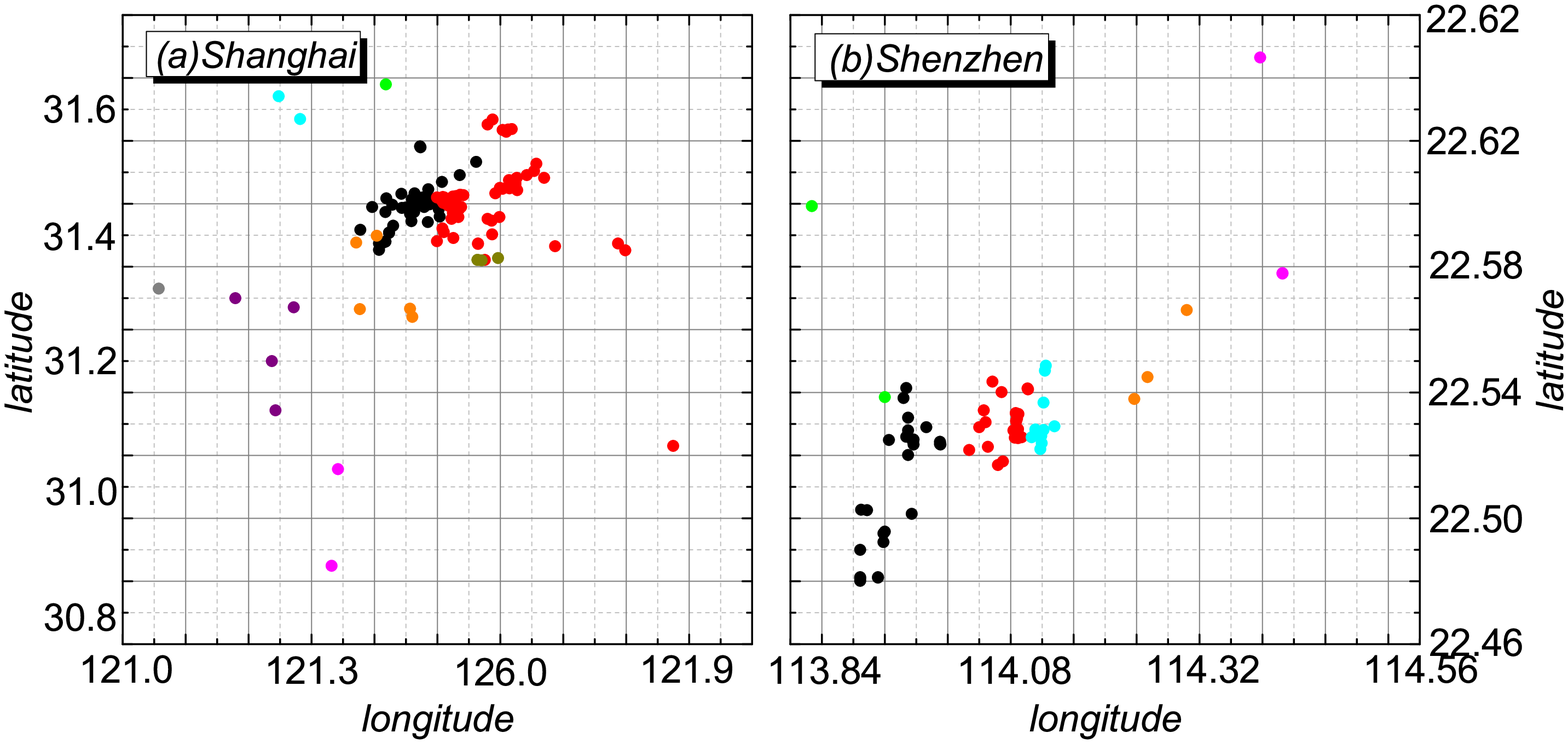}
\caption{The location distributions of the stocks inside the Shanghai and Shenzhen city are displayed in the panel (a) and (b), respectively. The stocks located in different districts of the city are marked by different colors. For the Shanghai city, the stocks located in pudong new district, baoshan district, jiading district, jinshan district, minhang district, nanhui district, songjiang district,qingpu district are marked by the red, green, cyan, magenta, orange, dark yellow, purple and gray circles, respectively. The huangpu, jingan, luwan, xuhui, changning, hongkou, yangpu, putuo and zhabei districts are combined to be marked by the black circles. For the Shenzhen city, the stocks located in Nanshan District, Baoan District, Futian District, Luohu District, Longgang district and Yantian District are marked by the black, green, red, cyan, magenta and orange circles, respectively.\label{f6}}
\end{figure}

\section*{Distance effect on stock correlation}

Internet shortens the distance of the world, and it becomes quite easy to access information. Then, should it change the geographical distance role in the stock markets, and should the long distance weaken the stock connection? To answer this question, we firstly investigate the stock distance distribution. The stock distance for the stock pairs inside each province is calculated. As shown in Fig. 5, the probability of the short distance is much greater than that of the long distance. A crossover behavior is found for the stock distance distribution with a two-stage decay, especially for the SH market. The crossover point is at about 87 kilometer (Km) for the SH market and 102 kilometer for the SZ market. The decay of the distribution can be fitted by the exponential function $f(x)=e^{a+bx}$, with $(a,b)=(-1.099,-0.022)$ and $(a,b)=(-2.251,-0.003)$ for the former and later stage of the SH market, and $(a,b)=(-1.648,-0.005)$ and $(a,b)=(-2.262,-0.003)$ for the former and later stage of the SZ market, respectively. The long distances should be mainly from the stock pairs located in different cities, and the inter-city distance is usually long. The short distances are mainly from the stock pairs inside the city. We then investigate the stock location distribution for several large cities, with figure 6 showing the stock location distribution of the Shanghai and Shenzhen city as an example. In Fig. 6, the stocks in different districts are marked by different colors. From the stock location distribution of the Shanghai and Shenzhen city, one may observe an inner-district clustering effect of stocks. The reason resulting in the clustering could be complicated. For a lot of Chinese cities, they usually have a city plan. For example, if a region of the city is planned to use as the industrial land, then a lot of companies would build the factories inside this region, which can partially explain the inner-district clustering of the stock location. Some other factors, such as the similar business, good infrastructures and transportation systems of the city, may also lead to the stock location clustering.

\begin{figure}[htb]
\centering
\includegraphics[width=10cm]{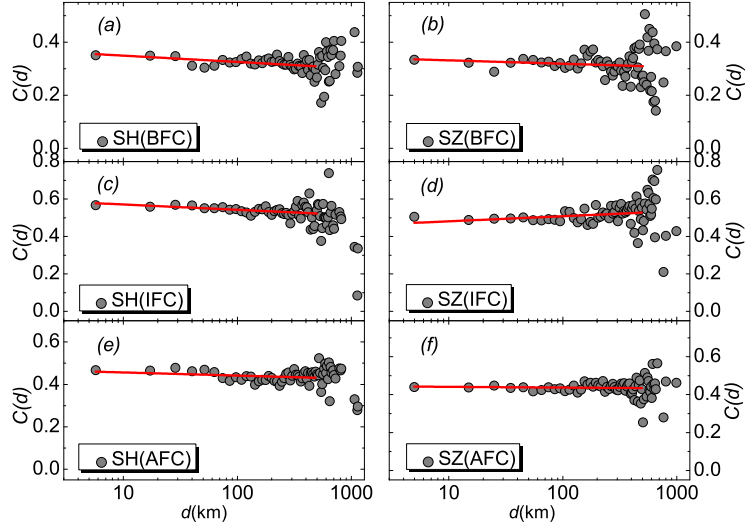}
\caption{The distance-dependent correlation $C(d)$ on the stock distance $d$ is displayed, with (a), (c), (e) for the periods before, in and after the financial crisis of the SH market, and (b), (d), (f) for the periods before, in and after the financial crisis of the SZ market, respectively. \label{f7}}
\end{figure}

\begin{figure}[htb]
\centering
\includegraphics[width=10cm]{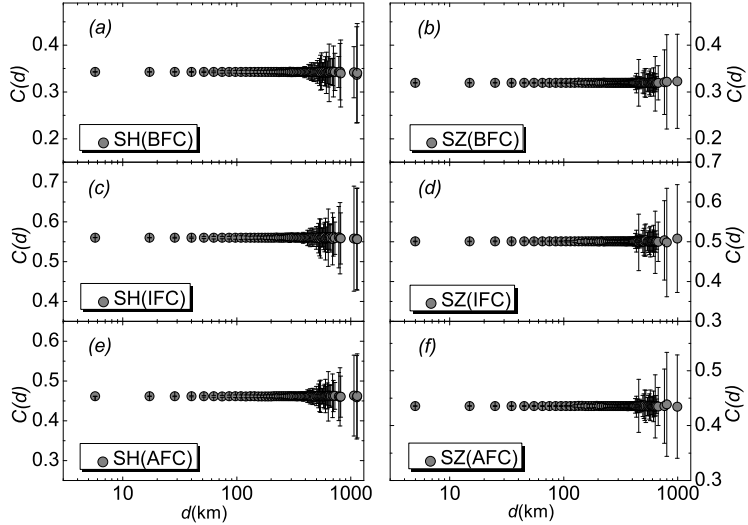}
\caption{A similar figure as Fig. 7, but for the shuffled data. The stems suggest the error bar. \label{f8}}
\end{figure}

Further, we study a distance-dependent correlation $C(d)$, which is defined as the average correlation for the stock pairs with the same distance,

\begin{equation}
{C(d)} = \langle c_{ij}|d \rangle
\end{equation}
where $d$ is the stock distance. Figure 7 shows the distance-dependent correlation $C(d)$ on the stock distance for the three periods, i.e., before, in and after the financial crisis, where the distance effect looks very weak, and a large fluctuation is found for the $C(d)$ on the long distance. This fluctuation can be explained by the sparse data of the long distance, which is observed in Fig. 5. To better understand the distance effect, we also perform a shuffling procedure for the data, and compute the average distance-dependent correlation $C(d)$ of the shuffled data over 1000 runs. As shown in Fig. 8, the $C(d)$ stays nearly unchanged for the shuffled data for all the three periods of the SH and SZ markets. To detect the difference between the real markets and the shuffled data, we do a t-test for the distance no more than 500 kilometers, to avoid the result inaccuracy induced by the large fluctuation of the long distances. The p-value of the t-test of the SH market and of the SZ market is displayed in table 1 and table 2, respectively. The t-test result manifests that under $1\%$ confidence level, the $C(d)$ of the SH market is significantly different from that of the shuffled data, while the $C(d)$ of the SZ market shows no significant difference from that of the shuffled data. By fitting the $C(d)$ on the stock distance, a weak decay is found for the SH market.

\begin{table*}
\caption{The p-value of the t-test for the SH market and the shuffled data.}
\begin{center}
\begin{tabular}{|c|c|c|c|c|}
\hline{Distance(Km)}
      & $\leq 200$ & $\leq 300$ & $\leq 400$ & $\leq 500$     \\
\hline{BFC} &$6.91\times10^{-4}$&$2.37\times10^{-5}$ & $1.80\times10^{-7}$& $5.32\times10^{-9}$\\
\hline{IFC} & $1.60 \times10^{-3}$ &$2.11\times10^{-5}$& $2.69\times10^{-5}$&$3.13\times10^{-5}$ \\
\hline{AFC} & $5.76\times10^{-4}$ & $2.27\times10^{-6}$ & $1.00\times10^{-7}$& $4.38\times10^{-9}$ \\
\hline
\end{tabular}
\end{center}
\end{table*}

\begin{table*}
\caption{The p-value of the t-test for the SZ market and the shuffled data.}
\begin{center}
\begin{tabular}{|c|c|c|c|c|}
\hline{Distance(Km)}
      & $\leq 200$ & $\leq 300$ & $\leq 400$ & $\leq 500$   \\
\hline{BFC} & $0.09$& $0.92$ & $0.20 $& $0.50$  \\
\hline{IFC}  &$0.73$ &  $0.21$ & $0.01$ & $0.02$ \\
\hline{AFC}&$0.49$ & $0.21$ & $0.11$& $0.89$ \\
\hline
\end{tabular}
\end{center}
\end{table*}

\section*{Conclusion}

The geography and distance effect on financial dynamics is investigated, based on the daily data of the individual stocks of the Chinese stock market. The financial center is found to still play an important role in stock dynamics, and the stocks located in the Shanghai city and the Guangdong province contribute greatly to the Shanghai and Shenzhen stock markets, respectively. The geographical correlation results show that the stock location has an influence on the financial dynamics, except for the financial crisis period of the Shenzhen stock market. The stock distance distribution is studied, and the probability of the short distance is observed to be much higher than that of the long distance. The Shanghai stock market shows a weak correlation decay with the stock distance, while the distance is found to have no impact on the stock correlation for the Shenzhen stock market.

\bigskip
{\textbf{Acknowledgments:}}

This work was partially supported by the National Natural Science Foundation of China (Grant Nos. 11175079, 71371165 and 11505099), the Jiangxi Provincial Young Scientist Training Project under Grant No. 20133BCB23017, and the Research Project of Zhejiang Social Sciences Association under Grant No. 2014N079.


\begin{thebibliography}{36}
\bibitem{man95} R. N. Mantegna, H. E. Stanley, {\it Nature} {\bf 376} (1995) 46.
\bibitem{gop99} P. Gopikrishnan, V. Plerou, L. A. N. Amaral,  H. E. Stanley, {\it Phys. Rev.} E {\bf 60} (1999) 5305.
\bibitem{liu99} Y. Liu,  P. Gopikrishnan, P. Cizeau, M. Meyer, C. K. Peng, H. E. Stanley, {\it Phys. Rev. E}  {\bf 60} (1999) 1390.
\bibitem{gu08} G.F. Gu, W. Chen, W.X. Zhou, {\it Physica A} {\bf 387} (2008) 495.
\bibitem{gab03} X. Gabaix, P. Gopikrishnan, V. Plerou,  H. E. Stanley, {\it Nature}  {\bf 423} (2003) 267.
\bibitem{ren10}  F. Ren, W. X. Zhou, {\it New J. of Phys.} {\bf 12} (2010) 075030.
\bibitem{mis10}  J. Miskiewicz, M. Ausloos, {\it Physica A} {\bf 389} (2010) 797.
\bibitem{zha11} L. Zhao,  G. Yang, W. Wang, Y. Chen, J. P. Huang, H. Ohashi, H. E. Stanley, {\it Proc. Natl. Acad. Sci. USA}
{\bf 108} (2011) 15058.
\bibitem{qiu12} T. Qiu, G. Chen, L. X. Zhong, X. R. Wu, {\it Physica A} {\bf 391} (2012) 2656.
\bibitem{pod09} B. Podobnik, D. Horvatic, A. M. Petersen, H. E. Stanley, {\it Proc. Natl. Acad. Sci. USA} {\bf 106} (2009) 22079.
\bibitem{ava14}  A. Avakian, B. Podobnik, M. Piskor, H. E. Stanley, {\it Phys Rev.} E {\bf 89} (2014) 032805.
\bibitem{cur14}  C. Curme, T. Preis, H. E. Stanley, H. S. Moat, {\it Proc. Natl. Acad. Sci. USA} {\bf 111} (2014) 11600.
\bibitem{ken15}  D. Y. Kenett, X. Huang, I. Vodenska, S. Havlin, H. E. Stanley, {\it Quantitative Finance} {\bf 15} (2015) 569.

\bibitem{qia15}  X. Y. Qian, Y. M. Liu, Z. Q. Jiang, B. Podobnik, W. X. Zhou, H. E. Stanley, {\it Phys. Rev.} E {\bf 91} (2015) 062816.

\bibitem{qiu08} T. Qiu , L. Guo, G. Chen, {\it Physica A} {\bf 387} (2008) 6812.
\bibitem{yam05} K. Yamasaki, L. Muchnik, S. Havlin, A. Bunde, H. E. Stanley {\it Proc. Natl. Acad. Sci. U.S.A} {\bf 102} (2005) 9424.
\bibitem{wan08} F. Z. Wang, K. Yamasaki, S. Havlin, H. E. Stanley {\it Phys. Rev. E} {\bf 77} (2008) 016109.
\bibitem{jun08} W. S. Jung, F. Z. Wang, S. Havlin, T. Kaizoji, H. T. Moon, H. E. Stanley {\it Eur. Phys. J. B} {\bf 62} (2008) 113.
\bibitem{gia01} I. Giardina, J. P. Bouchaud, M. Mezard, {\it Physica A} {\bf 299} (2001) 28.
\bibitem{son11} D. M. Song, M. Tumminello, W. X. Zhou, R. N. Mantegna, {\it Phys. Rev.} E {\bf 84} (2011) 026108.
\bibitem{men14} H. Meng, W. J. Xie, Z. Q. Jiang, B. Podobnik, W. X. Zhou, H. E. Stanley, {\it Scientific Reports } {\bf 4} (2014) 3566.

\bibitem{egu00} V. M. Eguiluz, M. G. Zimmermann, {\it Phys. Rev. Lett.} {\bf 85} (2000) 5659.
\bibitem{gu09} G. F. Gu, W. X. Zhou, {\it EPL} {\bf 86} (2009) 48002.
\bibitem{che13} J. J. Chen, B. Zheng, L. Tan, {\it PLoS ONE} {\bf 8} (2013) e79531.
\bibitem{qiu10} T. Qiu, B. Zheng, G. Chen, {\it New J. Phys.} {\bf 12} (2010) 043057.
\bibitem{hua15a} J. P. Huang, {Experimental Econophysics: Properties and Mechanisms of Laboratory Markets}, {\it Springer } (2015).
\bibitem{hua15b} J. P. Huang, {\it Physics Reports } {\bf 564} (2015) 1.

\bibitem{cas89} M. Castells, {The Informational City}, {\it Blackwell, Oxford } (1989).
\bibitem{ohm95a} K. Ohmae, {The Evolving Global Economy}, {\it Harvard Business Review Books, Cambridge, MA } (1995) .
\bibitem{ohm95b} K. Ohmae, {The End of the Nation-State: The Rise of Regional Economies}, {\it Harper Collins, London} (1995) .
\bibitem{kob97} S. J. Kobrin, {Electronic Cash and the End of National Markets}, {\it Foreign Policy (Summer)} (1997) 65-77.
\bibitem{ber97} B. J. L. Berry, E. C. Conkling, D. M. Ray, {The Global Economy in Transition, second ed.}, {\it Prentice Hall} (1997).
\bibitem{zha04} S. X. B. Zhao, L. Zhang, D. T. Wang, {\it Geoforum} {\bf 35} (2004) 577.
\bibitem{por05} R. Portes, H. Rey, {\it Journal of International Economics} {\bf 65} (2005) 269.
\bibitem{agg09} R. Aggarwal, C. Kearney, B. Lucey {\it Manuscript, FMA Annual Meeting} { } (2009).
\bibitem{luc10} B. M. Lucey, Q. Y. Zhang, {\it Emerging Markets Review} {\bf 11} (2010) 62.
\bibitem{pan07} R. K. Pan, S. Sinha, {\it Phys. Rev.} E {\bf 76} (2007) 046116.
\bibitem{she09} J. Shen, B. Zheng, {\it Europhys. Lett.} {\bf 86} (2009) 48005.
\bibitem{gop01} P. Gopikrishnan, B. Rosenow, V. Plerou, H. E. Stanley, {\it Phys. Rev.} E {\bf 64} (2001) 035106.
\bibitem{lal99} L. Laloux, P. Cizeau, J. P. Bouchaud, M. Potters, {\it Phys. Rev. Lett.} {\bf 83} (1999 ) 1467.
\bibitem{ple99} V. Plerou, P. Gopikrishnan, B. Rosenow, L. A. N. Amaral, H. E. Stanley,  {\it Phys. Rev. Lett.} {\bf 83} (1999) 1471.



\end{thebibliography}

\end{document}